\documentstyle[preprint,aps,epsf]{revtex}

\newcommand{\cmmt}[1]{{}}

\newcommand{\be}{\begin{equation}}
\newcommand{\ee}{\end{equation}}
\newcommand{\bear}{\begin{eqnarray}}
\newcommand{\eear}{\end{eqnarray}}
\newcommand{\ba}{\begin{array}}
\newcommand{\ea}{\end{array}}

\newdimen\tdim
\tdim=\unitlength
\def\bar{\overline}

\def\NPB#1#2#3{Nucl.~Phys.~B {\bf #1} (19#2) #3}
\def\PLB#1#2#3{Phys. Lett. B {\bf #1} (19#2) #3}

\def\PRD#1#2#3{Phys. Rev. D {\bf #1} (19#2) #3}

\tighten
\draft
\widetext
\begin{document}
\draft
\date{\today}
\title{
\normalsize
\mbox{ }\hspace{\fill}
\begin{minipage}{10 cm}
RUNHETC-2002-47,
UPR-1030-T
\\
{\tt hep-th/0303208}{\hfill}
\end{minipage}\\[5ex]
{\large\bf
Dynamical Supersymmetry
Breaking
in Standard-like Models
with Intersecting D6-branes
\\[1ex]}}

\author{Mirjam Cveti\v c $^{1}$\footnote{On Sabbatic Leave from University of
Pennsylvania.}, Paul Langacker $^{2}$, and Jing Wang$^{3}$
 \vspace{0.2cm}}
\address{$^1$ Department of Physics and Astronomy,\\ Rutgers University,
Piscataway, NJ 08855-0849, USA\\ and\\
School of Natural Science, Institute for Advanced Study,\\ Olden
Lane, Princeton,  NJ 08540, USA \\
$^2$ Department of Physics and
Astronomy, University of Pennsylvania, Philadelphia, PA 19104 \\
$^3$  Fermilab, Theory Division, Batavia, IL 60510, USA\\
}
\bigskip
\medskip
\maketitle

\def\kl#1{\left(#1\right)}
\def\th#1#2{\vartheta\bigl[{\textstyle{  #1 \atop #2}} \bigr] }

\begin{abstract}
{} We address dynamical supersymmetry breaking
  within a $N=1$ supersymmetric Standard-like Model based on a
   $Z_2\times Z_2$  Type IIA
 orientifold with intersecting D6-branes.
  The model possesses an additional, confining gauge sector  with the $USp(2)_A\times USp(2)_B
 \times USp(4)$ gauge group, where the  gaugino condensation mechanism  allows for the breaking of
  supersymmetry and stabilizes  moduli.
   We  derive  the  leading  contribution to the non-perturbative
  effective superpotential and determine numerically  the minima of the
    supergravity potential.
   These minima break supersymmetry  and  fix two undetermined moduli, which in turn
  completely specify the gauge couplings at the string scale. For this specific
   construction the minima have a negative cosmological constant.
  We expect that for  other supersymmetric Standard-like models with intersecting D6-branes,
   which also possess confining gauge sectors, the   supersymmetry
 breaking mechanism would have  qualitatively similar features.

\end{abstract}

\section{Introduction}

The second string revolution and the  advent of D-branes opened
the door for the  construction of open string solutions, which
correspond to the strongly coupled  heterotic string  sector. The
techniques of conformal field theory in describing D-branes and
orientifold planes  on orbifolds allow for the construction of
consistent
 four-dimensional $N=1$
supersymmetric models based on  Type II
orientifolds. Particular models, represented in
Refs.\cite{ABPSS,berkooz,N1orientifolds,zwart,ShiuTye,lpt,afiv,wlm,CPW,kr,CUW,aiqu},
are based on constructions with D-branes located at orbifold singularities, and chiral fermions
 appear   on the world-volumes of the
D-branes.

An alternative construction with chiral fermions, that has been explored only recently,
is that of Type II orientifold models with intersecting branes.
Chiral fermions appear in the open string spectrum,  localized at the intersections \cite{bdl}.
The model-building with intersecting branes
 was   developed \cite{bgkl,afiru,bkl,imr,magnetised}
(and subsequently explored in \cite{bonn,bklo,cim,bailin,kokorelis}),
where constructions of {\it non-supersymmetric} brane world
models were primarily addressed. Numerous examples
of non-supersymmetric three-family Standard-like models as well as
GUT models were obtained. However, the stability of
non-supersymmetric models is not well understood, especially when
the string scale is close to the Planck scale, since the
non-supersymmetric models  are subject to large quantum
corrections. Typically, the models are unstable when D-branes are
intersecting at angles, since supersymmetry is generically broken.

On the other hand,
examples of
${\cal N}=1$ supersymmetric orientifold models with branes at
angles were constructed
in \cite{CSU1,CSU2,CSU3},
 resulting in quasi-realistic models containing the
three-family  Standard Model. An example of a supersymmetric
$SU(5)$ GUT model with four families  of quarks and leptons
  (i.e., a net number of four ${\bf 10}$-plets and four ${\bf {\bar 5}}$-plets) was
also presented in \cite{CSU2}. The original construction is based
on  an ${\bf Z_2\times Z_2}$ orientifold with D6-branes  wrapping
specific supersymmetric three-cycles of the six-torus
($T^6=T^2\times T^2\times T^2$).

Recently, a  new example of the supersymmetric three-family
left-right symmetric model based on  an ${\bf T^6/Z_4}$ orientifold
was constructed \cite{blumrecent}.  Further  developments
\cite{CP} involve  the construction of  a larger class of
supersymmetric three-family Standard-like Models, based on ${\bf
T^6/(Z_2\times Z_2)}$ orientifolds,  by exploring the wrapping of
D6-branes along  more  general supersymmetric three-cycles (and
implementing RR tadpole cancellation conditions). A systematic
exploration of  a general class
 of  supersymmetric three-family $SU(5)$  GUT models  arising from ${\bf
T^6/(Z_2\times Z_2)}$ orientifolds with D6-branes wrapping
general supersymmetric three-cycles was most recently presented
in \cite{CPS}.

These quasi-realistic constructions provide a testing ground to
further address the phenomenology of such constructions.\footnote{
These models \cite{CSU1,CSU2,CSU3,blumrecent,CP,CPS} correspond in
the strong coupling limit to compactifications of M theory on
certain singular $G_2$ manifolds. As discussed in \cite{CSU3}, the
D-brane picture provides a description of how chiral fermions
arise from singularities of $G_2$ compactifications
\cite{AW,Witten,aW,CSU1,CSU2}. Recently, there has been an
exploration of phenomenological features  (e.g., the problem of
doublet-triplet splitting, threshold corrections, and proton
decay) of GUT models derived from $G_2$ compactifications
\cite{G2-23-split,Friedmann}. It would be interesting to explore
related features in this class of orientifold models.}
 A preliminary phenomenological study of  the first  three -family Standard-like
 model\cite{CSU1,CSU2}
was explored in \cite{CLS1,CLS2}.

In \cite{CLS1} a detailed study of the gauge couplings and their
renormalization group (RG) flow  were studied.
At the string scale these couplings depend on an additional modulus
parameter $\chi \equiv R^{1}_2/R^{1}_1$, where $R^{i}_{1,2}
$ are the respective radii of the $i$-th two-torus. The
Standard-Model gauge
 sector does not predict realistic low-energy values of gauge couplings
 (primarily due to the  additional Higgs and exotic fields in the massless
 spectrum). On the other hand  the  additional
non-Abelian gauge sector with the gauge group  $USp(2)_A\times
USp(2)_B\times Sp(4)$ has negative values of the $\beta$ functions,
and thus allows  for a confining phase in the infra red regime.
The  gaugino condensation can in turn take place and trigger
 dynamical supersymmetry breaking there.
 Charge confinement also implies the interesting feature that the left-handed members of
 an exotic ($SU(2)$-singlet) family  can become composite while their right-handed partners are
elementary.

 The main purpose of this note it to address dynamical
 supersymmetry breaking in the supersymmetric Standard-like model
 with intersecting D6-branes \cite{CSU1,CSU2}.
The  approach is is based on the study of  $N=1$ super Yang Mills
(SYM) theory with a confining phase in the infra red regime. There
the gaugino condensation generates a a non-perturbative
effective  superpotential \cite{VY}. A subsequent minimization of
the supergravity potential in turn determines the ground state,
which in certain cases breaks supersymmetry.
 (For recent exciting developments involving   the exact   non-perturbative
 superpotential, which includes all higher order instanton corrections,  for large classes
$N=1$ super Yang Mills theories, see \cite{DV,CDSW} and references therein.)

We shall  show that the additional gauge sectors of  the
supersymmetric Standard-like Model   allow for dynamical
supersymmetry breaking via gaugino condensation. For the specific
example we calculated the explicit dependence of the
non-perturbative superpotential on the moduli fields $S$ (dilaton)
and $U$ (complex structure modulus) of  one of the  three
internal two-tori; the other two are fixed due to the supersymmetry
constraint of the string construction. The minimization of the
explicit supergravity potential in turn produces isolated,
supersymmetry breaking
  minima, with both moduli $S$
and $U$ fixed.   These moduli  completely determine the
values of the gauge couplings in the theory at  the string scale.
Unfortunately, the specific example has the property that the
value of the potential at the  minimimum is negative and of the order of
the string scale.

While  we address a specific model,
 we expect that the qualitative features
 would be generic in other models,  with intersecting  branes
  and  confining  gauge sectors, such as constructed in \cite{CP}.
All of  these examples typically have a number of non-Abelian
confining  gauge group factors, typically associated with  $USp$
groups.  The non-perturbative superpotential, that is a
sum of exponential factors that  typically  depend on the
 dilaton $S$ and complex structure moduli $U_i$, will allow for minima in which
 such moduli are stabilized.

The paper is organized as follows. We  summarize
in Section 2 the results for the gauge group couplings and the explicit
 dependence of the gauge coupling on moduli $S$
and $U_i$, first in a general case of  models with intersecting
branes, and then for the specific model considered.
 In Section 3 we determine the explicit form of the non-perturbative
  superpotential, due to the leading instanton contribution,
  as a function of moduli and then focus on the
 concrete example.  We further minimize the supergravity potential numerically
 and analyse the features of the  minima, including implications for the values of gauge couplings and gaugino masses.
Conclusions are given in Section 4, where we contrast our results with
those for the  perturbative heterotic string constructions.


\section{Model}

\subsection{Essential Features of the Model}

In this section we shall provide the key features of the
construction. We refer the reader to the original
papers \cite{CSU1,CSU2} for more detailed discussions.

  For
concreteness, we consider an orientifold of type IIA on ${\bf
T}^6/({\bf Z}_2\times {\bf Z}_2)$. The orbifold actions have
generators $\theta$, $\omega$ acting as $ \theta: (z_1,z_2,z_3)
\to (-z_1,-z_2,z_3)$, and  $\omega: (z_1,z_2,z_3) \to
(z_1,-z_2,-z_3)$ on the complex coordinates $z_i$ of ${\bf T}^6$,
which is assumed to be factorizable. The orientifold action  is
$\Omega R$, where $\Omega$ is world-sheet parity, and $R$ acts by
$ R:\  (z_1,z_2,z_3) \to ({\overline z}_1,{\overline
z}_2,{\overline z}_3)$. The model contains four kinds of
O6-planes, associated to the actions of $\Omega R$, $\Omega
R\theta$, $\Omega R \omega$, $\Omega R\theta\omega$. The
cancellation of the RR crosscap tadpoles requires an introduction
of $K$ stacks of $N_a$ D6-branes ($a=1,\ldots, K$) wrapped on
three-cycles (taken to be the product of 1-cycles $(n_a^i,m_a^i)$
in the $i^{th}$ two-torus), and their images under $\Omega R$,
wrapped on cycles $(n_a^i,-m_a^i)$.

The cancellation of untwisted tadpoles imposes constraints on the
number of D6-branes and the types of 3-cycles that they wrap
around. The cancellation of twisted tadpoles determines the
orbifold actions on the Chan-Paton indices of the branes (which
are explicitly given in \cite{CSU1,CSU2}).

The condition that the system of branes preserves ${\cal N}=1$
supersymmetry requires \cite{bdl} that each stack of D6-branes is
related to the O6-planes by a rotation in $SU(3)$: denoting by
$\theta_i$ the angles the D6-brane forms with the horizontal
direction in the $i^{th}$ two-torus, supersymmetry preserving
configurations must satisfy $ \theta_1\, +\, \theta_2\, +\,
\theta_3\, =\, 0 $. This in turn impose a constraint on the
wrapping numbers and the complex structure moduli
$\chi_i=R_2^i/R_1^i$.

The rules to compute the spectrum are analogous to those in
\cite{bkl}. We summarize the resulting chiral spectrum in
Table \ref{matter}, found in \cite{CSU1,CSU2}, where
\begin{equation}
I_{ab}\ =\
(n_a^1m_b^1-m_a^1n_b^1)(n_a^2m_b^2-m_a^2n_b^2)(n_a^3m_b^3-m_a^3n_b^3)
\label{internumber}
\end{equation}
\begin{table} \footnotesize
\renewcommand{\arraystretch}{1.25}
\begin{center}
\begin{tabular}{|c|c|}
\hline {\bf Sector}   &
{\bf Representation} \\
\hline\hline
$aa$    &  \hspace{1cm} $U(N_a/2)$ vector multiplet \\
       & \hspace{1cm} 3 Adj. chiral multiplets  \\
\hline\hline $ab+ba$   &  $I_{ab}$ chiral multiplets in
$(N_a/2,\overline{N_b/2})$ rep.   \\
\hline\hline $ab'+b'a$ &  $I_{ab'}$ chiral multiplets in
$(N_a/2,N_b/2)$ rep.
  \\
\hline\hline $aa'+a'a$ &  $-\frac 12 (I_{aa'} - \frac{4}{2^{k}}
I_{a,O6})$
chiral multiplets in sym. rep. of $U(N_a/2)$  \\
          & $-\frac 12 (I_{aa'} + \frac{4}{2^{k}} I_{a,O6})$
chiral multiplets in antisym. rep. of $U(N_a/2)$ \\
\hline
\end{tabular}
\end{center}
\caption{\small General spectrum on D6-branes at generic angles
(namely, not parallel to any O6-plane in all three tori). The
spectrum is valid for tilted tori. The models may contain
additional non-chiral pieces in the $aa'$ sector and in $ab$,
$ab'$ sectors with zero intersection, if the relevant branes
overlap. \label{matter} }
\end{table}

 The D6-brane configuration for the  first  example leading to a three-family
Standard-like Model
is provided in table \ref{cycles3family}, and satisfies the
tadpole cancellation conditions. The configuration is
supersymmetric for
\be \chi_1:\chi_2:\chi_3=1:3:2\, .  \label{susyratio}  \ee
The weak hypercharge is given by
\be Y = (B-L)/2 + (Q_8+Q_{8'})/2,  \label{hypercharge}  \ee
where $B-L = Q_3/3 - Q_1$ and $Q_3$ is the charge
corresponding to the $U(1)$ in $U(3)_C$.

\begin{table}
\footnotesize
\renewcommand{\arraystretch}{1.25}
\begin{center}
\begin{tabular}{|c|c|c|l|}
\hline Type & Gauge Group &$N_a$ & $(n_a^1,m_a^1) \times
(n_a^2,m_a^2) \times (n_a^3,\widetilde{m}_a^3)$ \\
\hline
$A_1$ &$USp(8)\to U(1)_8\times U(1)_{8'}$& 8 & $(0,1)\times(0,-1)\times (2,{\widetilde 0})$ \\
$A_2$ & $USp(2)_A$&2 & $(1,0) \times(1,0) \times (2,{\widetilde 0})$ \\
\hline
$B_1$ &$U(2)_L$& 4 & $(1,0) \times (1,-1) \times (1,{\widetilde {3/2}})$ \\
$B_2$ &$USp(2)_B$ & 2 & $(1,0) \times (0,1) \times (0,{\widetilde {-1}})$ \\
\hline
$C_1$&$U(4)\to U(3)_C\times U(1)_{1}$ &6+2 & $(1,-1) \times (1,0) \times (1,{\widetilde{1/2}})$ \\
$C_2$ &$USp(4)$& 4 & $(0,1) \times (1,0) \times (0,{\widetilde{-1}})$ \\
\hline
\end{tabular}
\end{center}
\caption{\small D6-brane configuration for the three-family
model.} \label{cycles3family}
\end{table}

The resulting spectrum is given in the original paper \cite{CSU1,CSU2}
 and the subsequent papers  \cite{CLS1,CLS2}.

\subsection{Gauge Couplings}
\label{gauge} We shall summarize the results of the gauge
coupling calculations for the model. Since the gauge couplings are
associated with different stacks of branes they do not exhibit a
conventional gauge unification. Nevertheless, the value of each
gauge coupling at the string scale is predicted in terms of a
modulus $\chi$ and the ratio of the Planck to string scales. The
running is strongly affected by the exotic matter and multiple
Higgs fields, leading to low values of the MSSM sector couplings
at low energy. However, the hidden sector groups are
asymptotically free.

The gauge coupling of the gauge field from a stack of $D6$-branes
wrapping a three-cycle is given by
\begin{equation}
\frac{1}{g_{YM}^2} = \frac{ M_s^3 V_3}{(2 \pi)^4 g_s},\label{ymc}
\end{equation}
where $M_s=1/\sqrt{\alpha'}$ is the string scale, $g_s$ is the string coupling and $V_3$ is the volume of the
three-cycle wrapped by a particular D6-brane. For our specific cases $V_3$ is given
by\footnote{The definition of  $V_3$
in (\ref{v3}) differs from one in \cite{CLS1} by a factor of $1\over 4$.
This factor has to be included, due to
  the orbifolding of $T^6$ by ${\bf Z_2\times Z_2}$,
  which is an Abelian group of order 4. This implies that the
  expressions  in \cite{CLS1} for $g_{YM}^2$ and $\alpha_G$ should
  be  increased by a factor of 4. The numerical results in
  \cite{CLS1}, which were given for $M_s \sim M_P^{(4d)}$, are
  still approximately valid for the case $M_s \sim M_P^{(4d)}/4$.}:
\begin{equation}
V_3 = {\bf \textstyle{1\over 4}} (2 \pi)^3 \prod_{i=1}^3 \sqrt{n^{i2} \left(R_1^i\right)^2
+ \hat{m}^{i2} \left(R_2^i\right)^2}\, ,\label{v3}
\end{equation}
where $R_{1,2}^i$ are the  radii of the two dimensions of the
$i^{\rm th}$ two-torus, $\hat{m}^i = m^i$ for $i=1,2$, $\hat{m}^3
= \tilde{m}^3=m^3-\textstyle{1\over 2} n^3$ (the third $T^2$ is tilted), and the wrapping numbers $(n^i,\hat{m}^i)$ are
given in Table~\ref{cycles3family}.
One can trade $g_s$ in  (\ref{ymc}) for the four-dimensional Planck-scale  $M_{P}^{(4d)}$
which is defined as the coefficient of
the Einstein term in the low energy effective action:
\begin{equation}
S_{4d} = (M_P^{(4d)})^2 \int dx^4 \sqrt{g} R + \dots =
\frac{1}{16 \pi G_N} \int dx^4 \sqrt{g} R + \dots
\end{equation}
Since $G_N^{-1/2}=1.22 \times 10^{19}\ {\rm GeV}$, we have
$M_P^{(4d)} = \frac{1}{4 \sqrt{\pi}} \times G_N^{-1/2} = 1.7
\times 10^{18}$ GeV.
The Planck scale  is related
to the string coupling $g_s$  and string scale $M_s$ by
\begin{equation}
(M_P^{(4d)})^2 = \frac{ M_s^8 V_6 }{ (2 \pi)^7 g_s^2} \, ,\label{mpms}
\end{equation}
where $V_6$ is the total internal volume
given by
\begin{equation}
V_6 = \frac{ (2 \pi)^6}{4} \prod_{i=1}^3 R_1^i R_2^i.
\end{equation}
Again, the factor of $1\over 4$ is due to the orbifolding of
$T^6$ by ${\bf Z}_2 \times {\bf Z}_2$. This factor was included in \cite{CLS1}.
Employing (\ref{mpms}) allows us to write the gauge couplings in terms of $M_s$, $M_P^{(4d)}$,
$V_3$ and $V_6$:
\begin{equation}
g_{YM}^2=\sqrt{2\pi}{{M_s\sqrt{V_6}}\over{M_P^{(4d)}V_3}}\, ,
\end{equation}
which in terms  of the complex structure moduli $\chi_i = R_2^i/R_1^i$ becomes
\begin{equation}
g_{YM}^2 = \frac{\sqrt{8 \pi}M_s}{ M_P^{(4d)}}
\frac{\sqrt{\chi_1 \chi_2 \chi_3}}{ \prod_{i=1}^3 \sqrt{n^{i2} +
\hat{m}^{i2} \chi_i^2}}\, .
\end{equation}
The supersymmetric condition (\ref{susyratio}) implies
\begin{equation}\label{coupling}
g_{YM}^2 = \frac{4\sqrt{3 \pi} M_s}{M_P^{(4d)}}.
\frac{\chi^{3/2}}{\sqrt{[(n^1)^2 + (m^1)^2 \chi^2][ (n^2)^2 + 9
(m^2)^2 \chi^2][(n^3)^2 + 4 (\tilde{m}^3)^2 \chi^2]}},
\end{equation}
where $\chi \equiv \chi_1$.

At a scale $M$ below the string scale, the coupling $\alpha_a =
{{g_a^2}\over {4\pi}}$ of the $a^{\rm th}$ gauge factor is given (at one
loop) by
 \begin{equation}  \frac{1}{\alpha_a(M)} =
\frac{c_a(\chi)}{\alpha_G(\chi)} + \beta_a t, \label{rge1} \end{equation}
 where
\begin{equation} \alpha_G(\chi) = \sqrt{\frac{{3}}{{\pi}}} \frac{
M_s}{M_P^{(4d)}} \chi^{3/2}\, , \label{rge2}\end{equation}
 and
 \begin{equation}
t = \frac{1}{2\pi} \ln \frac{M_s}{M}. \label{rge3} \end{equation}
 The low
energy predictions for the model are given in \cite{CLS1} (There,
after correcting for the factor $1/4$ in (\ref{v3}),
$M_s \sim M_P^{(4d)}/4$ was assumed, and the low energy result
depends on one modulus parameter $\chi$.) Since we focus on the
additional confining gauge sector we state the values of $c_a$
and $\beta_a$ for these gauge couplings in Table~\ref{gaugefactors}.
In \cite{CLS1} the renormalization group  equations were studied without
the inclusion of the  chiral supermultiplets associated with the
open string sector of  the  brane.  There are three
copies of such states in the adjoint representation; they are due
to the fact that the supersymmetric cycles wrapped by D6-branes are not
rigid.
In the  Standard-model sector they     affect in a negative
way the low energy predictions for the standard model gauge couplings.
However, in the
quasi-hidden sector the only such states are associated with the $USp(4)$
gauge group, where they  change the beta function there from -5 to -2.
For  the sake of completeness we include them in the
study of gaugino condensation.

\begin{table} \footnotesize
\renewcommand{\arraystretch}{1.25}
\begin{center}
\begin{tabular}{|c||c|c|}
Group $a$& $c_a$ & $\beta_a({\rm int})$
\\
\hline
$USp(2)_B$  & $6\chi^2$  &  $-4$    \\
$USp(2)_A$  & 2  &  $-6$    \\
$USp(4) $  & $2 \chi^2$  &  $-2(-5)$  \\
\end{tabular}
\end{center}
\caption{\small Coefficients $c_a$ of $1/\alpha_G$, and $\beta$
functions $\beta_a$
for the  $USp(2)_B$ and $USp(2)_A$, associated with the $B_2$ and $A_2$ brane configuration,
respectively, and $USp(4)$  group associated with the $C_2$ brane configuration.
The beta function of $-2(-5)$ for $USp(4) $ includes (does not include) the
contributions of three chiral 5-plets that are not localized at intersections.
\label{gaugefactors}}
\end{table}

In the following subsection we shall derive the explicit complex
moduli dependence of the gauge couplings, which are suitable for
the determination of the effective non-perturbative superpotential.

\subsection{Gauge Kinetic Function and K\"ahler Potential
 in Terms of Complex Structure Moduli}

To determine the moduli dependence of the gauge
couplings in Type IIA theory with D6 branes in terms of complex
structure  moduli,  we shall employ the fact that
Type IIA theory with D6 branes is  T-dual to
 Type I theory with
D9 branes and background B-fluxes. Hence,   to arrive at
the proper definition of the moduli fields we shall  start by
writing down the moduli fields in the Type I theory with D9
branes, which  are well known (see for example \cite{Cremades}).
We then apply the duality transformations to arrive at the
moduli fields for D6 branes.

In the Type I string with D9 branes, the  real part of the  dilaton  $S$ has the
familiar expression:
\begin{equation}
Re(S)=  \frac{M_s^6 \Pi_{i=1}^{3} R_1^{i} R_2^{i}}{2\pi g_{s}}\, .
\end{equation}
 The  real part of the K\"ahler moduli  $T^i$ are
defined as
\be
Re(T^{i}) = \frac{M_s^2 R_{1}^{i} R_2^{i}}{2\pi g_{s}}.
\ee
Again, $R^{i}_{1,2}$  are the  radii of the of the $i$-th torus,
$M_s=\frac{1}{\alpha'}$ is the string scale ($\alpha'$ is the string tension) and $g_s$ is the
string coupling.

The T-duality transformations between D9 branes with B-fluxes and D6 branes
wrapped on 3-cycles are the following (see e.g., \cite{Cremades}),
\be
{R_2^{i}} \rightarrow \frac{1}{M_s^2 R_{2}^{i}},
\ee
\be
{R_1^{i}} \rightarrow R_1^{i},
\ee
\be
g_{s} \rightarrow \frac{g_{s}}{M_s^3\Pi_{i=1}^3 R_2^{i}}.
\ee

Under these transformations, the real part of the  dilaton
 $S$ and the complex structure moduli  $U^i$  take the following form:
\be
Re(S) = \frac{M_s^3 \Pi_{i=1}^3R_1^{i}}{2\pi g_{s}},
\ee
\be
Re(U^{i}) = \frac{M_s^3R_{1}^{i}R_{2}^{j}R_{2}^{k}}{2\pi g_{s}},
\ee
where $i\neq j\neq k$.

The expression for  $g_{YM}^2$ (\ref{ymc})  is determined  in terms of $V_3$ (\ref{v3}). The
supersymmetry constraint
for the particular model requires  the condition (\ref{susyratio}) on the
$\chi_i = \frac{R_2^{i}}{R_1^{i}}$.  It turns  out \cite{Cremades} that
these relations ensure that
the volume of the 3-cycle  $V_3$ in (\ref{v3}) can be written:
\begin{eqnarray}
V_3 &=& {\bf \textstyle{1\over 4}}(2\pi)^3(n^1n^2n^3 R_{1}^{(1)} R_{1}^{(2)}R_{1}^{(3)} -
n^1{\hat m}^2{\hat m}^3 R_{1}^{(1)} R_{2}^{(2)}R_{2}^{(3)} \cr
&-& {\hat m}^1n^2{\hat m}^3 R_{2}^{(1)} R_{1}^{(2)}R_{2}^{(3)} -
{\hat m}^1{\hat m}^2n^3 R_{2}^{(1)} R_{2}^{(2)}R_{1}^{(3)}) , \label{3vp}
\end{eqnarray}
where for the specific model ${\hat m}^i=m^i$ ($i=1,2$),
${\hat m}^3={\tilde m}^3= m^3-\textstyle{1\over 2} n^3$.
(It can be verified explicitly for  each set of  $(n^i, {\hat m}^i)$ in the model
 that eqs. (\ref{v3}) and (\ref{3vp}) are indeed equivalent.)

 Due to supersymmetry $g_{YM}^{-2}\equiv Re(f)$,
where the gauge kinetic function $f$ is a holomorphic function of the moduli $S$ and $U^i$.
Given the  above definition of the real part of the
 dilaton $S$ and the $U^i$ moduli and the form
of $g_{YM}^{-2}$ (\ref{ymc}), with the $V_3$
derived in (\ref{3vp}) one obtains:
\be
f =  {\bf \textstyle{1\over 4}} \left[ n^1n^2n^3 S -
n^1{\hat m}^2{\hat m}^3 U^1 -
{\hat m}^1n^2{\hat m^3} U^2 -
{\hat m}^1{\hat m}^2n^3 U^3 \right] \, .
\ee
It is indeed a holomorphic (and linear) function of the
 fields, as required by supersymmetry.

For the specific case of the additional  (quasi-hidden)gauge sector,
\be
f_{USp(4)} = \textstyle{1\over 4}U^2= \textstyle{1\over 12} U,
\ee
\be
f_{USp(2)_B} =  \textstyle{1\over 4}U^1= \textstyle{1\over 4} U,
\ee
\be
f_{USp(2)_A} =  \textstyle{1\over 2}S,
\ee
where the second equality in the above equations follows from (\ref{susyratio}),
which  implies:
\be
U^1 : U^2 : U^3 = 1: \frac{1}{3} : \frac{1}{2}.
\ee
 and $U\equiv U^1$.

For the sake of completeness we also quote the gauge  kinetic
functions  for the Standard Model sector:
\be
f_{[U(3)_C,U(1)_{1}]}= \textstyle{1\over 4}(S+\textstyle{1\over 2} U^2)= \textstyle{1\over 4}(S+\textstyle{1\over 6} U)\, ,
\label{sm1}\ee
\be
f_{U(2)_L}= \textstyle{1\over 4}(S+ \textstyle{3\over 2}U^1)=\textstyle{1\over 4}(S+ \textstyle{3\over 2}U)\,  ,
\label{sm2}\ee
\be
f_{[U(1)_8,U(1)_{8'}]}= \textstyle{1\over 2} U^3=\textstyle{1\over 4} U\, .
\label{sm3}\ee
From (\ref{sm1}), (\ref{sm3}), and (\ref{hypercharge}),
one finds
\be
f_{Y}= \textstyle{5\over {72}}(S+ \textstyle{{59}\over {30}}U)
\label{sm4}\ee
for weak hypercharge.

The {\it K\"ahler potential} for the  fields  is the
so called no-scale potential. It takes the following canonical form:
\be
K = -\log(S + \bar{S}) - \sum_{I=1}^3 \log(U^{I} +
\bar{U}^{I})= -\log(S+\bar{S}) - 3 \log(U + \bar{U}) + \log(6)\, .
\label{kahl}\ee
In (\ref{kahl}) and elsewhere we have set the Planck scale $M_P^{(4d)}$ to unity,
i.e., all dimensional quantities are scaled by appropriate powers of $M_P^{(4d)}$.

\section{Gaugino Condensate and Effective Moduli Potential}

In local supersymmetric theory  the gaugino condensate $\langle \lambda^{\alpha}\lambda_{\alpha}\rangle$
 is contained in the definition of a chiral  superfield:
\be
{\cal U} \equiv {\cal W}^\alpha {\cal W}_{\alpha},
\ee
where ${\cal W}_{\alpha}$ is the vector superfield whose fermionic component is $ \lambda^{\alpha}$.
The confinement scale $\mu$ of the strongly coupled theory is defined as the scale
at which the effective gauge coupling becomes strong and perturbation
theory breaks down. Consequently,  the gaugino condensate
$|\langle \lambda^{\alpha}\lambda_{\alpha}\rangle|\propto \mu^3$. This can be generated
by an exact effective superpotential for the chiral field ${\cal U}$~\cite{VY},
\be
W({\cal U}, \Phi) = \frac{1}{4}{\cal U}f_W(\Phi) - \frac{ {\cal U}}{32 \pi^2}
(\beta \log\frac{{\cal U}}{\Lambda^3} + const),
\ee
where $\Phi$ is  the modulus field in the theory which determines the strength
of the gauge coupling constants through the  gauge kinetic functions $f_W$;
$\beta$ is the $\beta$-function coefficient of the strongly coupled group, and
$\Lambda$ is the cut off scale of the theory.

The effective potential generates a VEV for ${\cal U}$,
\be
{\cal U} = \Lambda^3 \exp ( \frac{8\pi^2}{\beta} f_W(\Phi)) \times const.
\ee
Integrating out the field ${\cal U}$, an effective potential of the moduli
fields can be generated,
\be
W_{eff}(\Phi) = \frac{\beta}{32 \pi^2} \frac{\Lambda^3}{e}
\exp (\frac{8\pi^2}{\beta} f_W(\Phi)) \\
\equiv d \Lambda^3 \exp (bf_W(\Phi)),
\ee
where we have defined the  constants $d \equiv  \frac{\beta}{32e \pi^2}$ and
$b\equiv \frac{8\pi^2}{\beta}$.

The three groups $USp(4)$,
$USp(2)_A,$  and $USp(2)_B$  that  become strongly coupled have the beta function coefficients $\beta_1 = -2$,
$\beta_2 = -4$ and $\beta_3 = -6$, respectively. With the previously defined gauge kinetic
functions, the effective potential for the moduli fields $U$ and $S$ is
\be
W (U,S) = d_1 \Lambda^3 \exp (\frac{b_1}{12}U) +  d_2 \Lambda^3 \exp (\frac{b_2}{4}U) +
d_3 \Lambda^3 \exp ( \frac{b_3}{2}S),
\ee
where $d_i= \frac{\beta_i}{32e \pi^2}$ and $b_i =\frac{8\pi^2}{\beta_i}$.
This is of course only the leading instanton contribution to the
non-perturbative superpotential. It can be
justified post-factum if the negative exponents are large  at the
minimum of the potential. This indeed turns out to be
the case for the specific solution  discussed in the next Subsection.

\subsection{Scalar Potential and its Ground States}

Given the K\"ahler potential and the effective superpotential of the moduli
fields $U$ and $S$, one can derive the scalar potential for the moduli
fields,
\be
V= \frac{1}{(S + \bar{S}) (U+\bar{U})^3} \left\{ |(S+\bar{S})
\frac{\partial W}{\partial S} -W|^2 + 3|(\frac{U+\bar{U})}{3}
\frac{\partial W}{\partial U} -W|^2  -3 |W|^2 \right\}\, . \label{pot}
\ee
In the above potential we have absorbed   the coefficient $\sqrt{6}$
from the
$\log {6} $ term in the K\"ahler potential (\ref{kahl})
in  the
definition of $\Lambda^3\to \sqrt{6}\Lambda^3$.

It is expected that the  gauge coupling threshold corrections
would introduce  corrections that depend on toroidal   K\"ahler moduli
$T_i$ (see \cite{LS}) of the form that would modify the superpotential in
a multiplicative way, i.e.,
$W_{total}=W_0(T_i)W(S,U)$, where  $W_0$ typically depends on
a product of  Dedekind modular functions $\eta(T_i)$, i.e.,
$W_0\sim \prod_{i=1}^3 \eta(T_i)^{-2}$. A superpotential contribution of
that type, along with the
K\"ahler potential $K=-\prod_{i=1}^3 \log(T_i+{\bar T}_i)$, could in turn
also contribute to supersymmetry breaking    and   stabilization of
K\"ahler moduli.
In this paper we are not including these effects, i.e.,
we assume that the
dominant effects associated with the supersymmetry breaking come
from the tree level  gauge coupling contribution and are thus
associated with the $S$ and $U$ sector contributions. We hope to return to
the threshold correction contributions to the effective superpotential in
the future.

 It is difficult to derive an analytical expressions for the minimum of the potential, so
we proceed with a numerical analysis.

The potential is periodic in $Im(U)$ and $Im(S)$, with periods $12N/\pi$ and
 $3M/\pi$, respectively, where $M$ and $N$ are integers.  Thus, one can focus on finding the
value of  $Im(U)$ and $Im(S)$ in the  ``fundamental domain'' $\{0,12/\pi\}$ and $\{0,3/\pi\}$,
respectively.
The  numerical minimization
yields the minimum at: \be Re(S)=1.10, \ \ Re(U)=0.575,\ \
Im(S)=0.48+3M/\pi, \ \ Im(U)=1.91+12N/\pi \ee
Fig. 1 depicts the potential  near the
minimum as a
function of  moduli $S$  and $U$.

The value of the potential is negative at the minimum and is approximately
$  -3.56\ 10^{-3}\, \ L^2$, where $L\equiv \sqrt{6}\Lambda^3/(32\pi^2e)$. In the
potential (\ref{pot})  we have set the Planck scale  to 1.
The string scale is  typically chosen to be of the same
order as the Planck scale and thus
 $\Lambda ={\cal O}(1)$ (in Planck units). In our specific case
  (see the following subsection) $M_s\sim 1.85 M_P^{(4d)}$
 and thus $\Lambda^3\sim  6.33 [M_P^{(4d)}]^3 $. As a consequence $L\sim 1.81 \
 10^{-2}  [M_P^{(4d)}]^2$, and  the cosmological constant
 $\sim -1.16\times 10^{-6} [M_{P}^{(4d)}]^4$.
Since all the other
parameters of the potential  at the minima are fixed, the large
negative cosmological constant  is inevitable.

The terms  that dynamically break
supersymmetry are significantly smaller than the contribution
from  the $-3|W|^2$ term. In particular:
\be F_S K^{S {\bar S}}\bar F_{\bar{S}}= ( |(S+{\bar S})
\partial W / \partial S - W|^2/[(S+{\bar S})(U+{\bar U})^3]\sim 4.95\  10 ^{-7}
L^2\ , \ee
\be F_U K^{U{\bar U}}\bar F_{\bar{U}}=
 3|{{(U+{\bar U})}\over 3}
\partial W / \partial U - W|^2/[(S+{\bar S})(U+{\bar U})^3]\sim 1.03\ 10 ^{-7}
L^2\ , \ee
\be
 -3|W|^2 /[(S+{\bar S})(U+{\bar U})]^3\sim -3.56 \ 10^{-3} L^2\label{w3}\ .\ee
Here $F_\phi\equiv e^{K\over 2} (\partial_\phi W+ K_\phi W)$
and $K_\phi \equiv \partial_\phi K$.

In the case in which one does not include the matter
contribution that is associated with the open string sector of the
$USp(4)$ brane, its  beta function  changes from $-$2 to  $-$5. In the
latter case we  found  unstable points where $V\to -\infty$  as $Re(U)\to
0$.
One such  point corresponds to
$Im(U)=4.242,\ Re(S)=0.606,\ Im(S)=0.388$
This phenomenon  is due to  the fact that in this case the
negative contribution of the potential (arising from $-3|W|^2$
term) turns out to be dominant for small values of $Re(U)$.
We do not encounter this instability in the case $\beta=-2$, for which
the relative strengths of the exponents in the effective
superpotential balance in a way that the negative contributions to
the potential do not dominate for small values of $Re(U)$ and/or
$Re(S)$.

\begin{center}
\begin{figure}[ht]
\epsfysize=6cm
\epsfbox{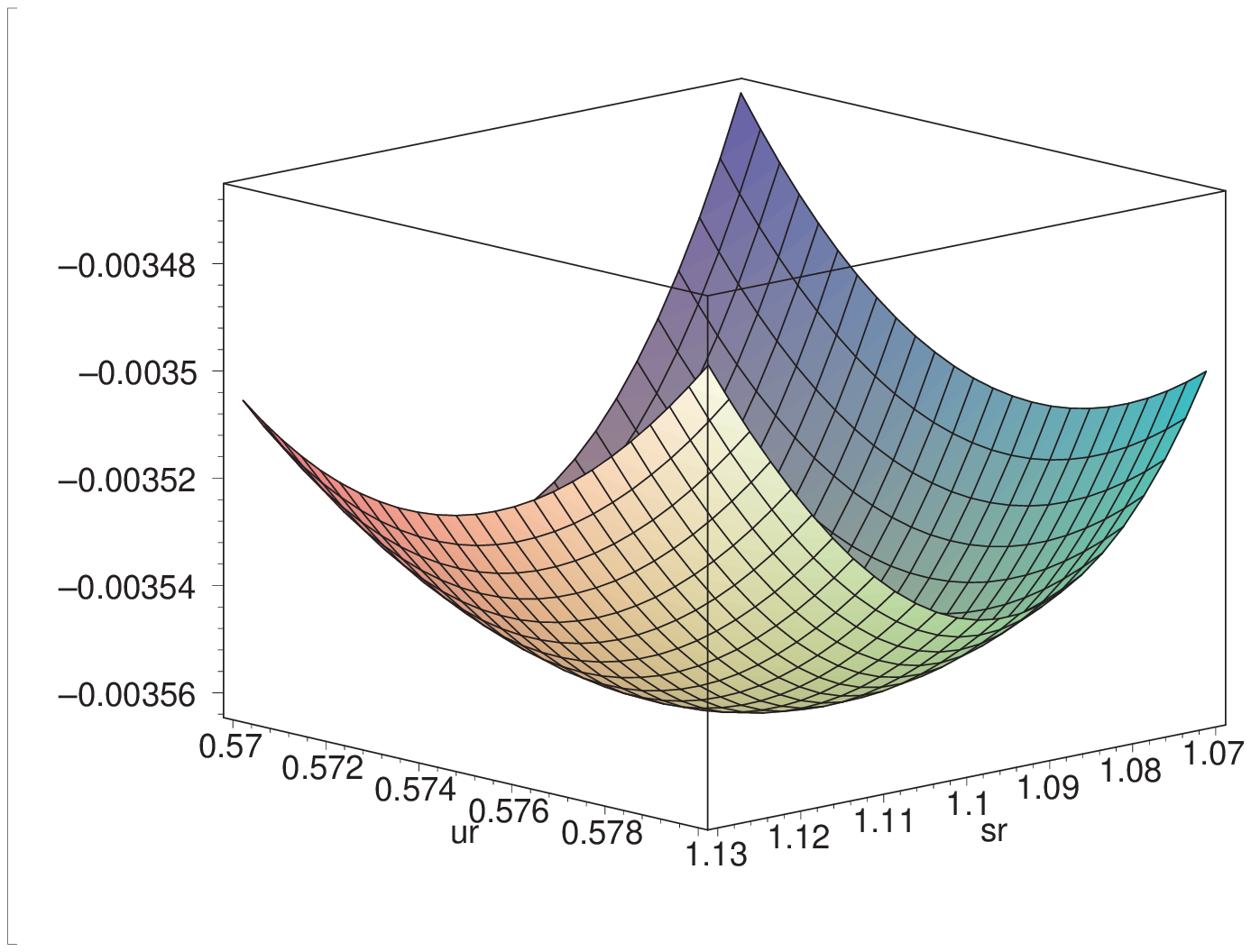}
\centering \epsfysize=6cm
\epsfbox{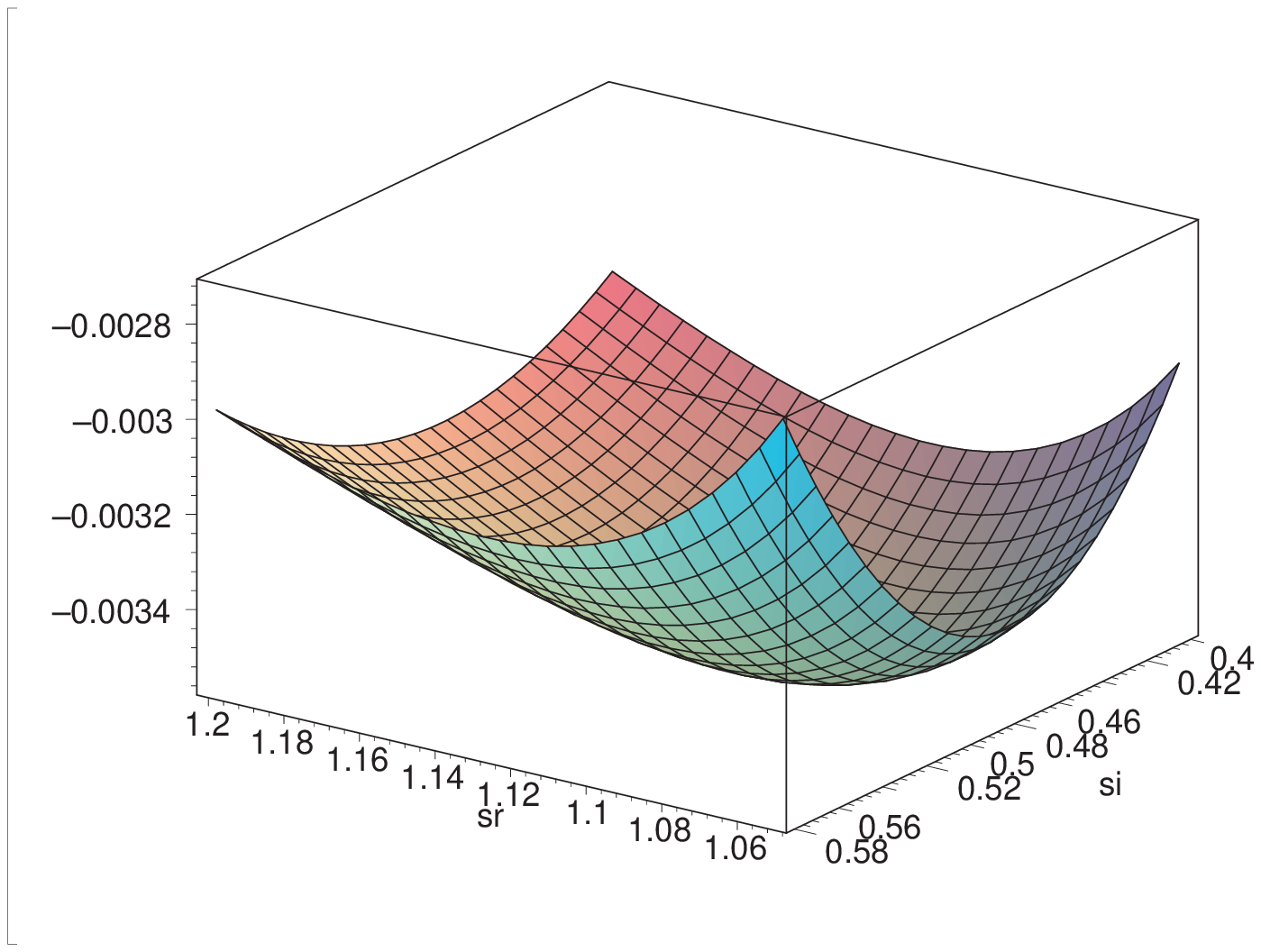}
\centering \epsfysize=6cm
\epsfbox{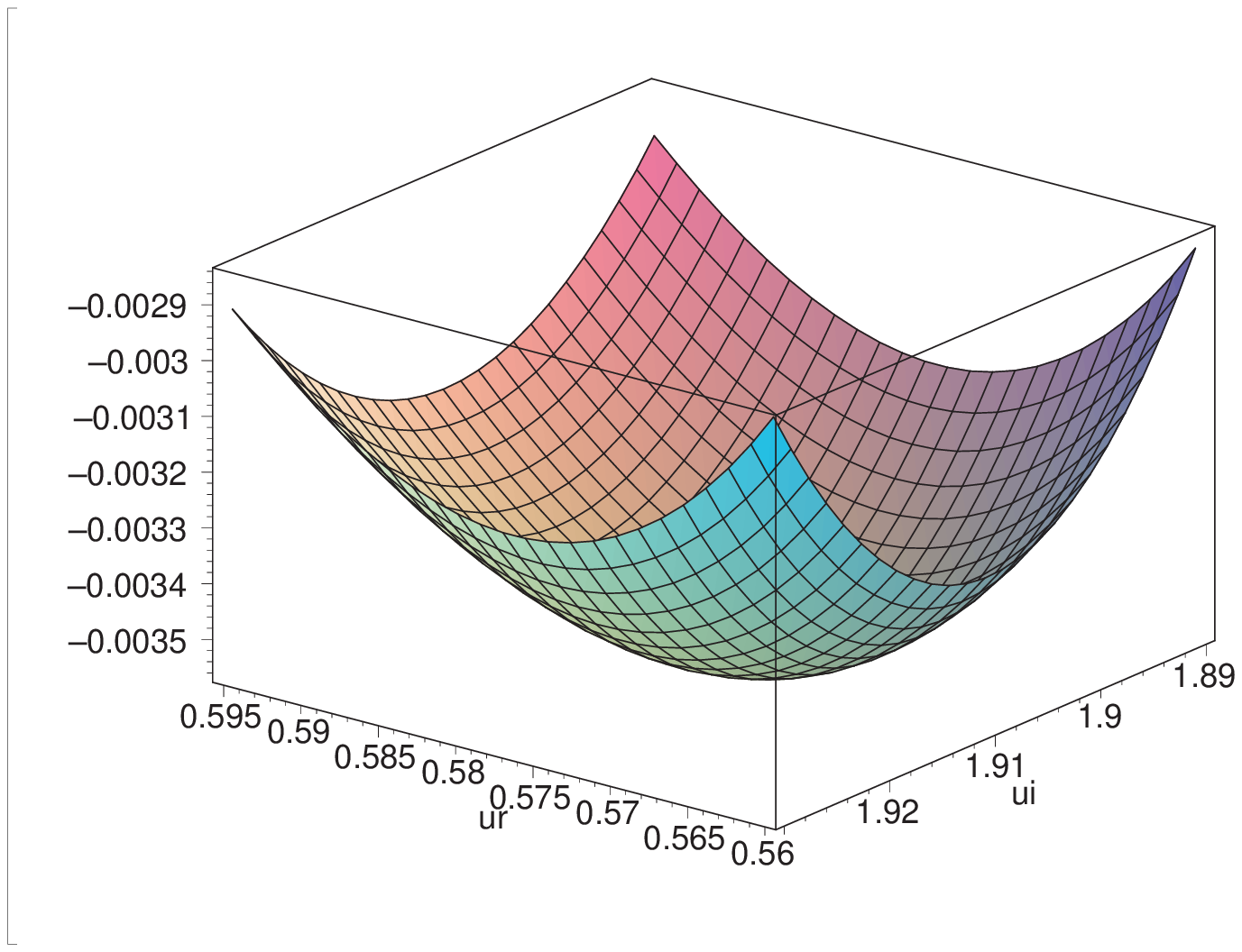}
\caption[]{\small Plots of  potential $V$  (in units of
$\Lambda^2\equiv [\sqrt{6}\Lambda^3/(32\pi^2e)]^2$) as a
function of $Re(S)$   and $Re(U)$
($Im(U)=1.91$ and $Im(S)=0.48$),
 $Re(S)$   and $Im(S)$
($Re(U)=0.575$ and $Im(U)=1.91 $), and
 $Re(U)$   and $Im(U)$
($Re(S)=1.10$ and $Im(S)=0.48$),  respectively.
In  the Figures $Re(S), \ Im(S), \ Re(U)$, and $Im(U)$ are denoted
$sr,\ si, \ ur, \ ui$, respectively.}
\label{orient}
\end{figure}
\end{center}
\vspace{2cm}

\subsection{Phenomenology}

We first  comment on features of the gauge couplings.
The quantities $\chi$ and $1/\alpha_G$ defined in (\ref{rge1}) and
(\ref{rge2}) are related to the moduli by
\be
\chi  =  \sqrt{\frac{Re (U)}{6 Re (S)}},  \ \ \ \
\frac{1}{\alpha_G}  =  \pi Re (S)\, . \ee
At the minimum of the potential they take the values
$\chi = 0.295$ and $1/\alpha_G = 3.46$, corresponding to
$ M_P^{(4d)}/M_s = 0.542$, which satisfies the
perturbative consistency condition $ M_P^{(4d)}/M_s > 1/\sqrt 8 \pi$~\cite{CLS1}.
From these values, and the expressions for
the MSSM gauge parameters $c_a$ and $\beta_a$ given in Table VI
of~\cite{CLS1}, we can calculate the predicted values of the
standard model gauge couplings at the electroweak scale\footnote{A fully
realistic construction would predict the electroweak scale from the
soft supersymmetry breaking. In our case, we simply use the
experimental electroweak scale, which corresponds to $t=6.06$ in (\ref{rge3}).}.
The inverse strong and electromagnetic
couplings are predicted to be
\be
\frac{1}{\alpha_3} = 52.2, \ \ \ \
\frac{1}{\alpha} = 525\, , \ee
which are much larger than the respective  experimental values $\sim 8.5$ and
128.  The  unrealistically small values predicted for the gauge couplings
are due to the extra chiral matter in the construction\footnote{Unlike in~\cite{CLS1},
we are also including the chiral states that are not localized at the
brane intersections for consistency with our treatment of the strongly coupled
confining sector. With these states, the strong $SU(3) $ group
is not asymptotically free.}.
The weak angle $\sin^2 \theta_W$, which is a ratio of gauge couplings,
fares somewhat better: it is predicted to be 0.29, not
too far from the experimental 0.23.

Unfortunately, since the  minima have negative cosmological
constants, these vacua do not provide  realistic backgrounds for
a  detailed study of the soft supersymmetry breaking parameters
of the charged matter sector of the model.
We defer this investigation for the future.

We can however determine gaugino masses in terms of the $F_S$ and $F_U$.
The general expression for the gaugino mass  (i.e., terms  of the type
$\lambda_a\lambda_a$  in the Lagrangian), is
\be
m_{\lambda_a}=(\partial_{\phi_i}\,f_a)K^{\phi_i \, {\bar\phi_j}}
{\bar F_{\bar \phi_j}}\, .
\ee
Here $K^{\phi_i \, {\bar\phi_j}}$ is the inverse of the K\"ahler metric,
$f_a$ is a gauge kinetic function,
and $F_\phi$ was defined after Eq. (\ref{w3}) .
 In the Standard Model sector the  gauge functions, determined
  in (\ref{sm1}-\ref{sm3}),
yield  the following  expressions for
 gaugino masses at the string scale:
\begin{eqnarray}
m_{[U(3)_C,U(1)_{1}]}&=&\textstyle{1\over 4}K^{S\, {\bar S}}{\bar F_{\bar S}}
+ \textstyle{1\over {24}}K^{U\,
{\bar U}}{\bar F_{\bar U}}\\ \nonumber
&=&  (1.89-3.48\, I)\, 10^{-4}\, L\sim (3.42-6.30\, I)\, 10^{-6}
M_P^{(4d)}\,,
\end{eqnarray}
\be
m_{U(2)_L}=\textstyle{1\over 4}K^{S\, {\bar S}}{\bar F_{\bar S}}+
 \textstyle{3\over {8}}K^{U\,
{\bar U}}{\bar F_{\bar U}}=(2.41-3.95\, I)\, 10^{-4}
 L \sim  (4.36-7.15\, I)\, 10^{-6} M_P^{(4d)}\, ,
\ee
\be
m_{[U(1)_8,U(1)_{8'}]}=\textstyle{1\over 4}K^{U\,
{\bar U}}{\bar F_{\bar U}}= (3.93-3.56\, I)\,10^{-5} L\sim
(7.11-6.44\, I)
10^{-7} M_P^{(4d)}\, ,
\ee
where we have restored the appropriate factor of $M_P^{(4d)}$ in the final expressions.
When a set of $U(1)$'s with charges $Q_a$ is broken to a single $U(1)'$
with charge $Q' = \sum_a d_a Q_a$, then the $U(1)'$ coupling and
gaugino masses are related to those of the original factors by
\be
\frac{1}{\alpha'} = \sum_a \frac{d_a^2}{\alpha_a}, \ \ \ \
m' =  \frac{\sum_a \frac{d_a^2}{\alpha_a} m_a}{ \sum_a \frac{d_a^2}{\alpha_a}}\, . \ee
From (\ref{hypercharge}) one then obtains\footnote{In the present case, the additional
$U(1)$ factors are not broken at a high scale. $m_Y$ therefore refers to
the diagonal $YY$ element of the gaugino mass matrix.}
\be
m_{Y}= (1.20 -2.03\, I)\,10^{-4}\, L\sim (2.17-3.67\, I) \,
10^{-6} M_P^{(4d)}
\ee

These masses are non-universal, complex (indicating significant $CP$-violating phases),
and  the values for the specific solution  are too large.
As in all such constructions, the gaugino masses below the string scale satisfy the
same RG equations at one loop as the corresponding gauge couplings, so that
$m_a(t)/m_a(0) = \alpha_a(t)/\alpha_a(0)$. However, unlike
heterotic constructions and simple grand unified theories, the gaugino
masses and gauge couplings at the string scale depend on two moduli $S$ and $U$.
These dependences are non-universal and are different for the gaugino masses and
gauge couplings. Thus the  gaugino unification prediction
$m_b(t)/m_a(t) = \alpha_b(t)/\alpha_a(t)$ of those models is lost.
Rather, one has
\be
\frac{m_b(t)\alpha_a(t)}{m_a(t)\alpha_b(t)} = \frac{m_b(0) f_b}{m_a(0) f_a}\, .
\label{nonuniversal}\ee
For example, for the minimum of the potential in this model, the
right hand side of (\ref{nonuniversal}) is $0.52-0.026 I$ for $(b,a) =
(SU(3),SU(2))$ and $10.6+ 0.16 I$ for $(b,a) =
(SU(2),\sqrt{\frac{5}{3}} U(1)_Y)$,
where $\sqrt{\frac{5}{3}} U(1)_Y$
corresponds to the coupling $5 \alpha_Y/3$ that unifies with $\alpha_2$ and
$\alpha_3$ in the conventional MSSM.

 \section{Conclusions}

 We  conclude with a few remarks contrasting the results obtained  with those of the
 perturbative heterotic quasi-realistic models.
 The supersymmetry breaking   in  heterotic  models has been extensively studied
 (see, e.g., \cite{Casas,Casasp} and references therein, and for recent studies \cite{Abel}.).
 One specific feature of heterotic models is that the tree level gauge couplings are universal
 and depend only on one modulus $S$. Therefore,  the gaugino
  condensation typically  generates an effective superpotential that
  involves only one field, thus making the minimization of the supergravity potential
  a more intricate process. In addition, for
  a number of quasi-realistic models, while possessing an
  additional gauge sector, such sectors  often were not confining (the beta functions were
  positive due to the additional matter).  Further exploration involved  the  string
  threshold corrections that depend on toroidal moduli  and allow for additional
  features of the  supersymmetry breaking  vacuum. In  these
  examples the cosmological constant was in general large and
  negative and would have to be fixed  by hand.

  In contrast the supersymmetric  models with intersecting D6 branes provide a framework with a
  confining gauge sector, where gaugino condensation can be addressed explicitly. We
  have demonstrated in   an explicit example that the effective non-perturbative superpotential
  allows for the minimum of the supergravity potential in which  supersymmetry is broken
  and the moduli (that determine the tree level gauge couplings at the string scale) are
  completely determined. Since the gauge couplings  typically  depend on more than one modulus,
  the minimization of the potential involves an interplay among all these moduli. The specific
  example has the property that the part of the potential that spontaneously breaks
   supersymmetry  is much smaller than the  $-3|W|^2$ term, resulting in a large  and negative
   cosmological constant. We hope that other quasi-realistic models \cite{CP}
   with intersecting
   D6 branes  may
   remedy this feature and possibly yield more realistic predictions,
   and plan to investigate dynamical
   supersymmetry breaking there.

\section*{Acknowledgments}

We would like to thank  F. Cachazo, M. Douglas and  expecially G. Shiu for
useful discussions. M.C. would like to thank the  New Center for
Theoretical Physics at Rutgers University  and the Institute for
Advanced Study, Princeton,
 for hospitality and
support  during the course of this work.
The research was supported in part by DOE grant DOE-EY-76-02-3071,
 NATO
the linkage  grant No. 97061 (M.C.), and the Fay R. and Eugene L. Langberg
Chair (M.C.).

\end{document}